\documentclass[10pt]{iopart}

\usepackage{graphicx}
\usepackage{dcolumn}
\usepackage{bm}
\usepackage{color}

\begin{document}

\title[CFETR]{Theoretical study of the Alfven Eigenmode stability in CFETR steady state discharges}


\author{J. Varela}
\ead{jvrodrig@fis.uc3m.es}
\address{Universidad Carlos III de Madrid, 28911 Leganes, Madrid, Spain}
\author{J. Huang}
\ead{juan.huang@ipp.ac.cn}
\address{Institute of Plasma Physics, Chinese Academy of Sciences, Hefei, Anhui, China}
\author{D. A. Spong}
\address{Oak Ridge National Laboratory, Oak Ridge, Tennessee 37831-8071, USA}
\author{J. Chen}
\address{Institute of Plasma Physics, Chinese Academy of Sciences, Hefei, Anhui, China}
\author{V. Chan}
\address{University of Science and Technology of China, Hefei, China}
\author{L. Garcia}
\address{Universidad Carlos III de Madrid, 28911 Leganes, Madrid, Spain}
\author{A. Wingen}
\address{Oak Ridge National Laboratory, Oak Ridge, Tennessee 37831-8071, USA}
\author{Y. Ghai}
\address{Oak Ridge National Laboratory, Oak Ridge, Tennessee 37831-8071, USA}
\author{Y. Zou}
\address{Southwestern Institute of Physics, Chengdu 610041, China}

\date{\today}

\begin{abstract}
The aim of this study is to analyze the stability of Alfven Eigenmodes (AE) in the China Fusion Engineering Test Reactor (CFETR) plasma for steady state operations. The analysis is done using the gyro-fluid code FAR3d including the effect of the acoustic modes, EP Finite Larmor radius damping effects and multiple energetic particle populations. Two high poloidal $\beta$ scenarios are studied with respect to the location of the internal transport barrier (ITB) at $r/a \approx 0.45$ (case A) and $r/a \approx 0.6$ (case B). Both operation scenarios show a narrow TAE gap between the inner-middle plasma region and a wide EAE gap all along the plasma radius. The AE stability of CFETR plasmas improves if the ITB is located inwards, case A, showing AEs with lower growth rates with respect to the case B. The AEs growth rate is smaller in the case A because the modes are located in the inner-middle plasma region where the stabilizing effect of the magnetic shear is stronger with respect to the case B. Multiple EP populations effects (NBI driven EP + alpha articles) are negligible for the case A, although the simulations for the case B show a stabilizing effect of the NBI EP on the $n=1$ BAE caused by $\alpha$ particles during the thermalization process. If the FLR damping effects are included in the simulations, the growth rate of the EAE/NAE decreases up to $70 \%$, particularly for $n > 3$ toroidal families. Low $n$ AEs ($n<6$) show the largest growth rates. On the other hand, high $n$ modes ($n=6$ to $15$) are triggered in the frequency range of the NAE, strongly damped by the FLR effects.
\end{abstract}

%
%
%
%
%


\vspace{2pc}

\maketitle

\ioptwocol

\section{Introduction \label{sec:introduction}}

The CFETR device has an important role in the pathway to the industrial exploitation of the nuclear fusion energy \cite{1}, required to test the technology and operation scenarios developed in the ITER project \cite{2} toward the construction of a DEMO reactor \cite{3}. The main target of CFETR is to demonstrate the feasibility of generating energy outputs between 200 MW to 1 GW, sustaining a high duty factor and a tritium breeding factor above unity in steady state operation.

The steady state operation scenario is proposed for ITER \cite{4,5,6} and CFETR \cite{7,8,9} plasmas, explored in Tokamaks \cite{10} as DIII-D \cite{11,12}, EAST \cite{13,14}, ASDEX \cite{15,16}, JET \cite{17} and KSTAR \cite{18}. Among the different steady state operation scenarios, the high poloidal $\beta$ discharges performed in JT60U \cite{19}, DIII-D \cite{20,21,22} and EAST \cite{23,24} devices are encouraging candidates for ITER and CFETR. High poloidal $\beta$ discharges show a large bootstrap current fraction, required in non inductive operations \cite{25,26}, and a high edge safety factor that reduces the possibility of disruptions \cite{27}. In addition, these discharges have an improved MHD instability (second stability regime), favorable transport properties, a high confinement factor as well as reactor extrapolations with reasonable device size and fusion output power. DIII-D/EAST joint experiments exploring high poloidal $\beta$ scenarios show the formation of internal transport barriers (ITB) leading to an improved energy confinement and Greenwald fractions close to unity \cite{21,28}. 

A CFETR high poloidal $\beta$ discharge with efficient plasma heating requires the minimization or avoidance of Alfven Eigenmode (AE) instabilities triggered by energetic particles (EP). Energetic particle driven instabilities cause an enhancement of the transport of fusion produced alpha particles, energetic hydrogen neutral beams and particles heated using ion cyclotron resonance heating (ICRF) \cite{29,30}, thus a fraction of these particles populations are lost before thermalization. Such performance deterioration is observed in several devices such as DIII-D and EAST tokamaks or LHD and W7-X stellarators \cite{31,32,33,34} if the mode frequency resonates with the drift, bounce or transit frequencies of the EP. Experimental and theoretical studies of high poloidal $\beta$ discharges in DIII-D and EAST show the destabilization of AEs \cite{35,36,37,38}, thus AEs could be also triggered in CFETR steady state operations. Consequently, a detailed analysis of the AE stability in CFETR steady state scenarios is mandatory to identify optimized operational regimes.

CFETR is a Tokamak with a major radius of $7.2$ m and a minor radius of $2.2$ m, an elongation $\kappa = 2$, a magnetic field intensity at the magnetic axis of $6.5$ T and a plasma current up to $14$ MA. CFETR plasma is heated by a combination of NBI, electron cyclotron waves (ECW) and low hybrid waves (LHW). The projected tangential negative-ion-based neutral beam (N-NBI) will deposit Deuterium particles with an energy of $350$ keV at the magnetic axis, injecting a power of $5$ MW. 

The goal of the present study is to analyze the stability of the AEs triggered by the NBI driven EP and alpha particles in CFETR steady state scenarios. Several resonances during the thermalization process of the NBI driven EP and alpha particles (EP energies) for different population densities (EP $\beta$) are considered. This information is useful to identify the destabilization threshold of the AEs with respect to the EP intensity during the slowing down process, as required for optimization studies. Two different configurations are analyzed with respect to the location of the internal transport barrier (ITB), at $r/a \approx 0.45$ (case A) and at $r/a \approx 0.6$ (case B). In addition, the plasma AE stability is studied for the individual destabilizing effect of the NBI driven EP and Alpha particles as well as the combined effect of both EP populations, including the stabilizing effect of the finite Larmor radius (FLR) on the EP. For that purpose, a set of simulations are performed using the FAR3d code \cite{39,40,41,42}.

This paper is organized as follows. An introduction to the numerical scheme is presented in section \ref{sec:model}. First, the study of the low $n$ AE stability in CFETR steady state discharges with respect to the destabilizing effect of the NBI EP and $\alpha$ particles individually is shown in section \ref{sec:Single}. Next, in section \ref{sec:Multiple}, the low $n$ AE stability of the CFETR steady state scenario is analyzed with respect to multiple EP populations including the FLR effects on the EP and thermal ions. The analysis of the high $n$ AE stability is performed in section \ref{sec:High}. Finally, the conclusions of this paper are presented in section \ref{sec:conclusions}.

\section{Numerical scheme \label{sec:model}}

The FAR3d code solves the linear reduced, resistive MHD equations for the thermal plasma (poloidal flux, total pressure, toroidal component of the vorticity and thermal parallel velocity) coupled with the equations of the EP density and parallel velocity moments \cite{43,44}, adding the linear wave-particle resonance effects required for Landau damping/growth and the parallel momentum response of the thermal plasma required for coupling to the geodesic acoustic waves \cite{45}. The code variables evolve using a set of equilibria calculated by the VMEC code \cite{46}. The numerical model uses finite differences in the radial direction and Fourier expansions in the angular variables for the equilibrium flux coordinates ($\rho$, $\theta$, $\zeta$). The coefficients of the closure relation are selected to match analytic Toroidal AE (TAE) growth rates based upon a two-pole approximation of the plasma dispersion function (Maxwellian EP distribution). 

It should be noted that a single EP Maxwellian distribution and a slowing down distribution do not induce the same AE stability because the gradient of the phase space distribution determines the drive of the AE modes. We expect that these effects can in the future be incorporated into a gyro-Landau closure model. This remains a topic for future research. This analysis limitation is mitigated by performing a parametric analysis with respect to the EP energy and $\beta$. In this matter, the resonances triggered by a slowing down distribution function are estimated using a set of Maxwellian distribution functions. The present model reproduces the destabilizing effect of the passing EP, although the AEs triggered by ICRF driven EP or anisotropic beams cannot be modeled correctly. However, the tangential NBIs in CFETR plasma generate EPs with small pitch angles thus the model approximation is valid.

An eigenvalue solver is used to resolve the linear equation set, providing the growth rate and frequency of the dominant (mode with the largest growth rate) and sub-dominant modes. The analysis of the sub-dominant modes indicate the growth rate of the multiple AE families that can be unstable or marginally unstable during CFETR discharges.

The model was applied in the AE stability analysis of high poloidal $\beta$ and reverse shear discharges in DIII-D \cite{38,47}, LHD \cite{48} and TJ-II \cite{49} plasma, indicating reasonable agreement with the observations. 

\subsection{Equilibrium properties}

\begin{figure}[h!]
\centering
\includegraphics[width=0.5\textwidth]{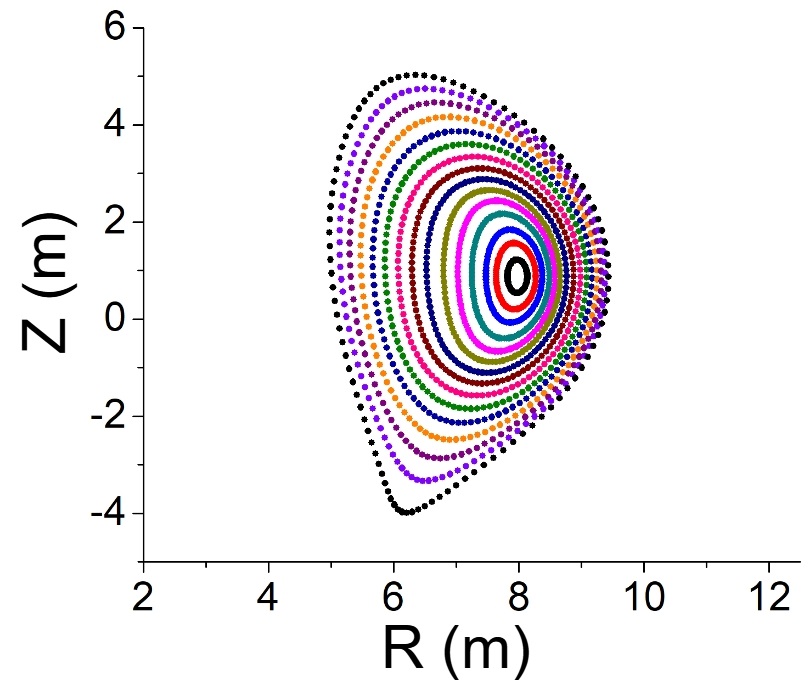}
\caption{Magnetic surfaces of model A calculated by VMEC code}\label{FIG:1}
\end{figure}

Two fixed boundary results from the VMEC equilibrium code \cite{46} are used for CFETR steady state configurations maintaining $1$ GW of fusion power, a plasma current of $11$ MA, a Greenwald density fraction of $1.1$ and a bootstrap fraction of $71 \%$. The VMEC equilibria are transformed from EFIT equilibria \cite{50} calculated using the core-pedestal coupled integrated modelling work flow in OMFIT \cite{51}. NBI EP and alpha particle density and energy profiles are calculated by ONETWO/NUBEAM \cite{52,53} and TRANSP/NUBEAM \cite{54} simulations. The case A corresponds to a hypothetical discharge with the ITB located at $r/a \approx 0.45$ and the case B at $r/a \approx 0.6$ \cite{55}. Figure~\ref{FIG:1} shows the magnetic surfaces of the case A (similar to case B).

\begin{figure}[h!]
\centering
\includegraphics[width=0.5\textwidth]{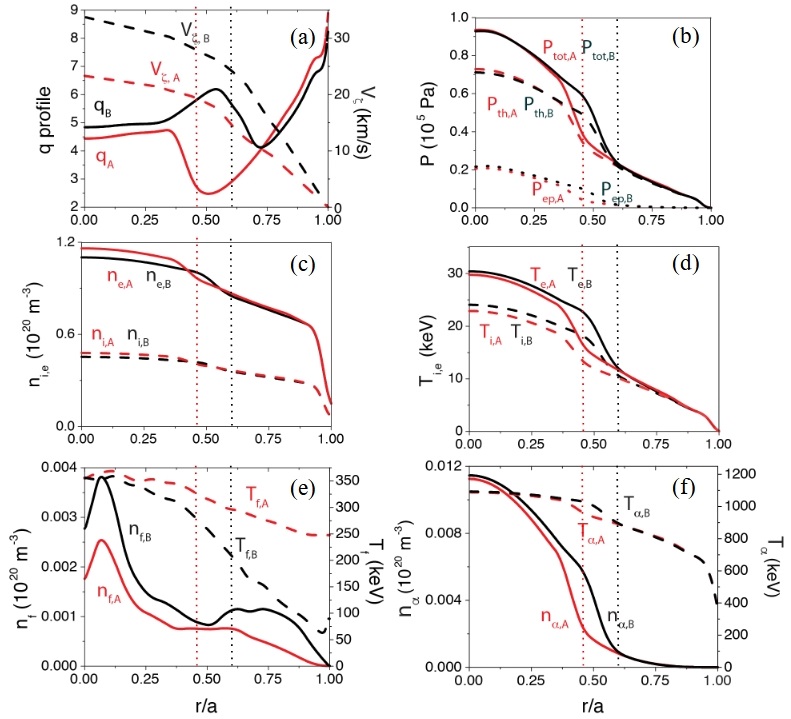}
\caption{Main profiles of the cases A (red lines) and B (black lines). (a) q profile (left Y axis, solid lines) and equilibrium toroidal plasma rotation (right Y axis, dashed lines), (b) Total (solid lines), thermal (dashed lines) and EP (dotted lines) pressure, (c) thermal electron (solid lines) and ion (dashed lines) density, (d) thermal electron (solid lines) and ion (dashed lines) temperature, (e) density (left Y axis, solid lines) and energy (right Y axis, dashed lines) of the NBI driven EP and (f) density (left Y axis, solid lines) and energy (right Y axis, dashed lines) of the $\alpha$ particles. The dotted vertical lines indicate the location of the IT in case A (red) and case B (black).}\label{FIG:2}
\end{figure}

Figure~\ref{FIG:2} indicates the main profiles of the cases A and B. The panel (a) shows the q profile and equilibrium toroidal plasma rotation, the panel (b) the total and EP pressure, the panel (c) the thermal electron and ion density, the panel (d) the thermal electron and ion temperature, the panel (e) the density and energy of the NBI driven EP and the panel (f) the density and energy of the Alpha particles. The location of the ITB in both configurations are linked to a plasma region with strong magnetic shear (large decrease of the safety factor), a sharp decrease of the equilibrium toroidal rotation as well as a flattening of the thermal and EP pressure profiles. Consequently, the location of the ITB may lead to a configuration with different stability features. The magnetic field at the magnetic axis is $6.5$ T and the averaged inverse aspect ratio is $\varepsilon=0.305$. A Deuterium NBI and a D-T plasma with $M_{ion} = 2.5$ are assumed. The model indicates $Z_{eff} \approx 2$. The effect of the impurities are not included directly in the analysis for simplicity.

\begin{figure}[h!]
\centering
\includegraphics[width=0.35\textwidth]{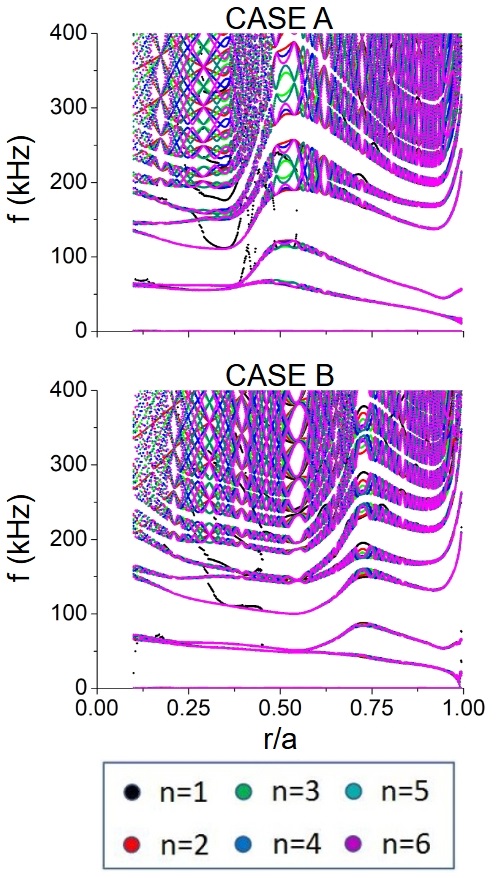}
\caption{Continuum gaps of cases A and B for the $n=1$ to $6$ toroidal mode numbers.}\label{FIG:3}
\end{figure}

Figure~\ref{FIG:3} shows the continuum gaps of the cases A and B for the $n=1$ to $6$ toroidal mode families calculated using the Stellgap code \cite{56} (The toroidal family indicates the set of poloidal modes that share the same toroidal number, that is to say, poloidal modes are coupled due to the poloidal variation of the magnetic field in the tokamak). The upper frequency range of the Beta induced AE gap (BAE) is $60$ kHz in the case A except in the middle-outer plasma region where upper frequency increases to $90$ kHz. On the other hand, the BAE gap upper frequency in case B is $50$ kHz all across the normalized minor radius. There is a narrow Toroidal AE (TAE) gap in the inner plasma region for the case A reaching the upper frequency range ($140$ kHz) in the middle-outer plasma region. The case B also shows a narrow TAE gap between the inner and middle plasma region and the upper frequency limit ($100$ kHz) is observed at the plasma periphery. The TAE gaps are narrow due to the large magnetic shear in the reverse shear region as well as the relatively flat safety factor, thermal ion density and thermal electron temperature in the inner plasma region. For both cases there is a wide Elliptic AE (EAE) gap between $60 - 115$ kHz in the inner plasma region, reaching the middle plasma for the case B. Also, there are several Non circular AE (NAE gaps) at higher frequencies. It should be noted that the continuum gaps of case B are similar to case A although displaced radially outward by $\Delta \approx 0.2$. Some discontinuities are observed in the $n=1$ continuum bands near the magnetic axis, originated by the VMEC transformation to Boozer coordinates in the inner plasma region.

\subsection{Simulations parameters}

The dynamic toroidal modes ($n$) in the simulations range from $n=1$ to $15$ and the dynamic poloidal modes ($m$) are selected to cover the main resonant rational surfaces excluding the plasma periphery ($r/a > 0.85$). The dynamic modes evolve during the simulation and the equilibrium modes ($n=0$) do not evolve and represent the equilibria. The safety factor profiles change between case A an B thus the mode selection is different. Table~\ref{Table:1} shows the equilibrium and dynamic modes for the cases A and B.

\begin{table}[t]
\centering
\begin{tabular}{c c c}
\hline
n & m (case A) & m (case B) \\ \hline
0 & [0,10] & [0,10] \\
1 & [2,6] & [4,7] \\
2 & [5,10] & [8,13] \\
3 & [7,15] & [12,20] \\
4 & [10,20] & [16,26] \\
5 & [12,25] & [20,33] \\
6 & [15,30] & [24,39] \\
7 & [17,35] & [28,45] \\
8 & [20,40] & [32,52] \\
9 & [22,45] & [36,58] \\
10 & [25,50] & [40,65] \\
11 & [27,55] & [44,71] \\
12 & [30,60] & [48,78] \\
13 & [32,65] & [52,84] \\
14 & [35,70] & [56,91] \\
15 & [38,75] & [60,97] \\ \hline
\end{tabular}
\caption{Equilibrium and dynamic modes in the simulations. The first column indicates the toroidal modes, the second columns the poloidal modes in the case A and the third column the poloidal modes in the case B. Both parities are included for equilibrium and dynamic modes.} \label{Table:1}
\end{table}

The closure of the kinetic moment equations breaks the MHD parities so both parities must be included for all the dynamic variables. In addition, up-down asymmetric equilibria also require sine and cosine components. The convention of the code with respect to the Fourier decomposition is, in the case of the pressure eigenfunction, that $n>0$ corresponds to $cos(m\theta + n \zeta)$ and $n<0$ corresponds to $sin(m \theta + n \zeta)$. For example, the Fourier component for $2/1$ mode is $cos(2 \theta + 1 \zeta)$ and for the mode $-2/-1$ is $sin(2 \theta + 1 \zeta)$. The mode eigenfunction is represented using the following Fourier expansion:
\begin{eqnarray} 
f(\rho, \theta, \zeta, t) = \sum_{n,m} f^{s}_{mn} (\rho, t) sin(m\theta + n\zeta) \nonumber\\
+ \sum_{n,m} f^{c}_{mn} (\rho, t) cos(m\theta + n\zeta)
 \end{eqnarray}
with $f^{s}_{mn}$ and $f^{c}_{mn}$ real functions for the sine and cosine components, respectively. The magnetic Lundquist number is assumed $S=5\cdot 10^6$. The number of radial points is $100$.

Different NBI EP and $\alpha$ particle resonances are analyzed during the EP thermalization process (EP energies). Particularly, we include in the study two different groups of $\alpha$ particles that correspond to the $\alpha$ particle population with an energy of $1090$ keV ($\alpha$ type I), as well as the $\alpha$ particle population with energies of $800$ and $600$ keV ($\alpha$ type II). It should be noted that the birth energy of the $\alpha$ particles is $3.5$ MeV, thus the study also includes the energies $2000$ and $3000$ keV, although the analysis of the type I $\alpha$ is focused on $1090$ keV because this is the averaged energy of Helium 4 particles in the model, that is to say, the population of 1090 keV $\alpha$ particles is the largest leading to a stronger destabilizing effect compared to $\alpha$ particles in the range of $3.5$ to $1$ MeV. Likewise, the NBI EP population with an energy of $350$ keV corresponds to the NBI EP type I and the NBI EP population with the energies $275$, $200$ and $125$ keV to the NBI EP type II. The type I populations indicate the resonance caused by NBI EP with the peak energy of the NBI and $\alpha$ particles with the averaged Helium 4 energy. Type II populations shows resonances triggered by NBI EP during the thermalization process and $\alpha$ particles with an energy below the averaged Helium 4 energy. In addition, the destabilization threshold of the AEs is studied performing simulations with different $\alpha$ and NBI EP $\beta$ values.

\section{Simulations with a single EP population for low $n$ AEs \label{sec:Single}}

First, the AE stability in CFETR steady state scenarios is analyzed with respect to the individual destabilizing effect of the EP injected by the NBI and due to the $\alpha$ particles. The simulations are performed for different NBI EP and $\alpha$ particles energies, so that different resonances along the slowing down process are considered in the study. In addition, for each EP energy a set of EP $\beta$ are analyzed identifying the EP $\beta$ threshold for the AE destabilization. The reference model $\beta$ at the magnetic axis of the NBI EP ($\beta_{f}$) is $0.00135$ for the case A and $0.00257$ for the case B. Regarding the $\alpha$ particles, the $\beta$ at the magnetic axis ($\beta_{\alpha}$) is $0.04667$ for the case A and $0.04756$ for the case B.

\subsection{Case A: ITB in the middle plasma region}

\begin{figure*}[h!]
\centering
\includegraphics[width=0.75\textwidth]{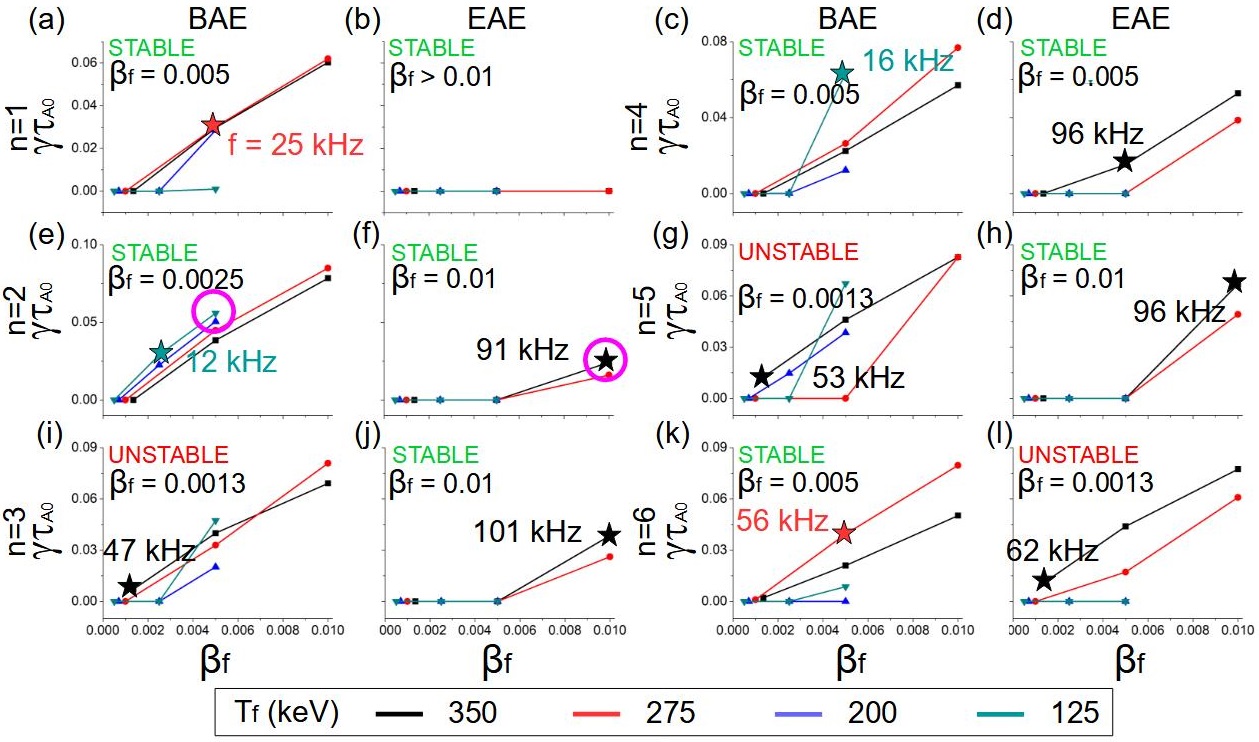}
\caption{Growth rate of $n=1$ to $6$ dominant modes for different EP energies and $\beta$ values destabilized by NBI driven EP. The green/red capital letter in the top of the panel indicates if the mode is stable/unstable for the reference operational regime of the NBI. The minimum NBI EP $\beta$ required to destabilize AE is also indicated. The colored stars indicate the unstable mode with the largest growth rate and lower EP $\beta$ threshold, including the mode frequency. The pink circles indicate the modes for which their eigenfunction is plotted in the fig. 6.}\label{FIG:4}
\end{figure*}

Figure~\ref{FIG:4} shows the growth rate and frequency of the $n=1$ to $6$ dominant modes (modes with the largest growth rate) for different energies and $\beta$ value destabilized by the NBI driven EP. The simulations indicate stable AEs for the reference NBI operational regime except for the $n=3$ and $n=5$ BAEs with $47$ and $53$ kHz, respectively, as well as the $n=6$ EAE with $62$ kHz, showing relatively small growth rates ($\gamma \tau_{A0} \approx 0.01$). Nevertheless, if the EP $\beta$ increases to $0.005$, type II NBI EPs lead to the destabilization of $n=1$ to $n=6$ BAEs and $n=6$ EAE. If the EP $\beta$ further increases to $0.01$, $n=2$ to $n=5$ EAEs are also unstable.

\begin{figure}[h!]
\centering
\includegraphics[width=0.5\textwidth]{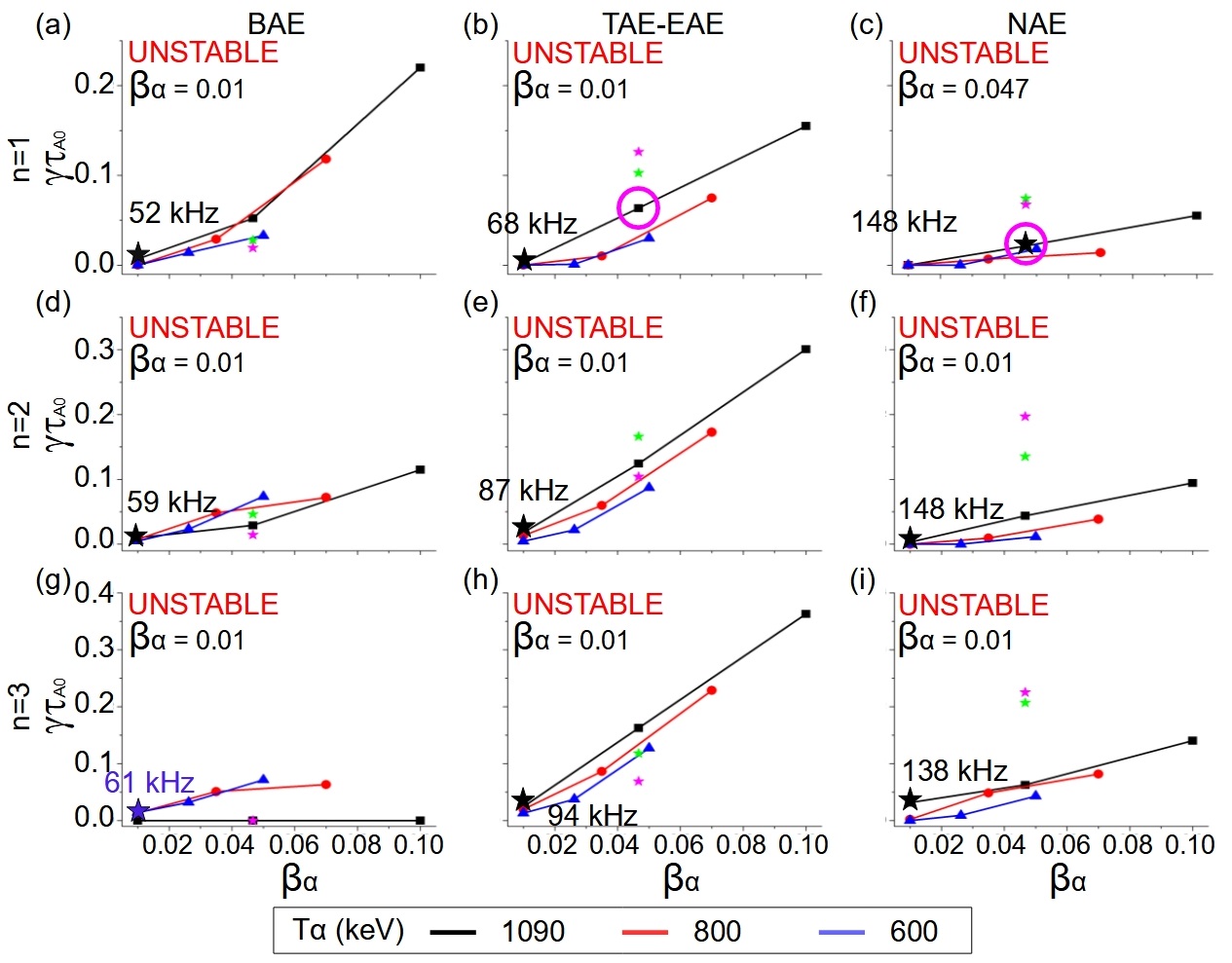}
\caption{Growth rate of $n=1$ to $3$ dominant modes for different EP energies and $\beta$ values destabilized by the $\alpha$ particles. The green/red capital letter in the top of the panel indicate if the mode is stable/unstable for the expected $\alpha$ particle population in the reference model. The colored stars indicate the modes with the largest growth rate and lower $\beta_{\alpha}$ threshold, including the mode frequency. The pink circles indicate the modes for which their eigenfunction is plotted in fig. 6. The $\beta_{\alpha}$ required to destabilize AE is also indicated. The green (pink) star indicates the growth rate of the AEs destabilized by $\alpha$ particles with $T_{\alpha}=2000$ ($3000$) keV and $\beta_{\alpha} = 0.04667$.}\label{FIG:5}
\end{figure}

\begin{figure}[h!]
\centering
\includegraphics[width=0.5\textwidth]{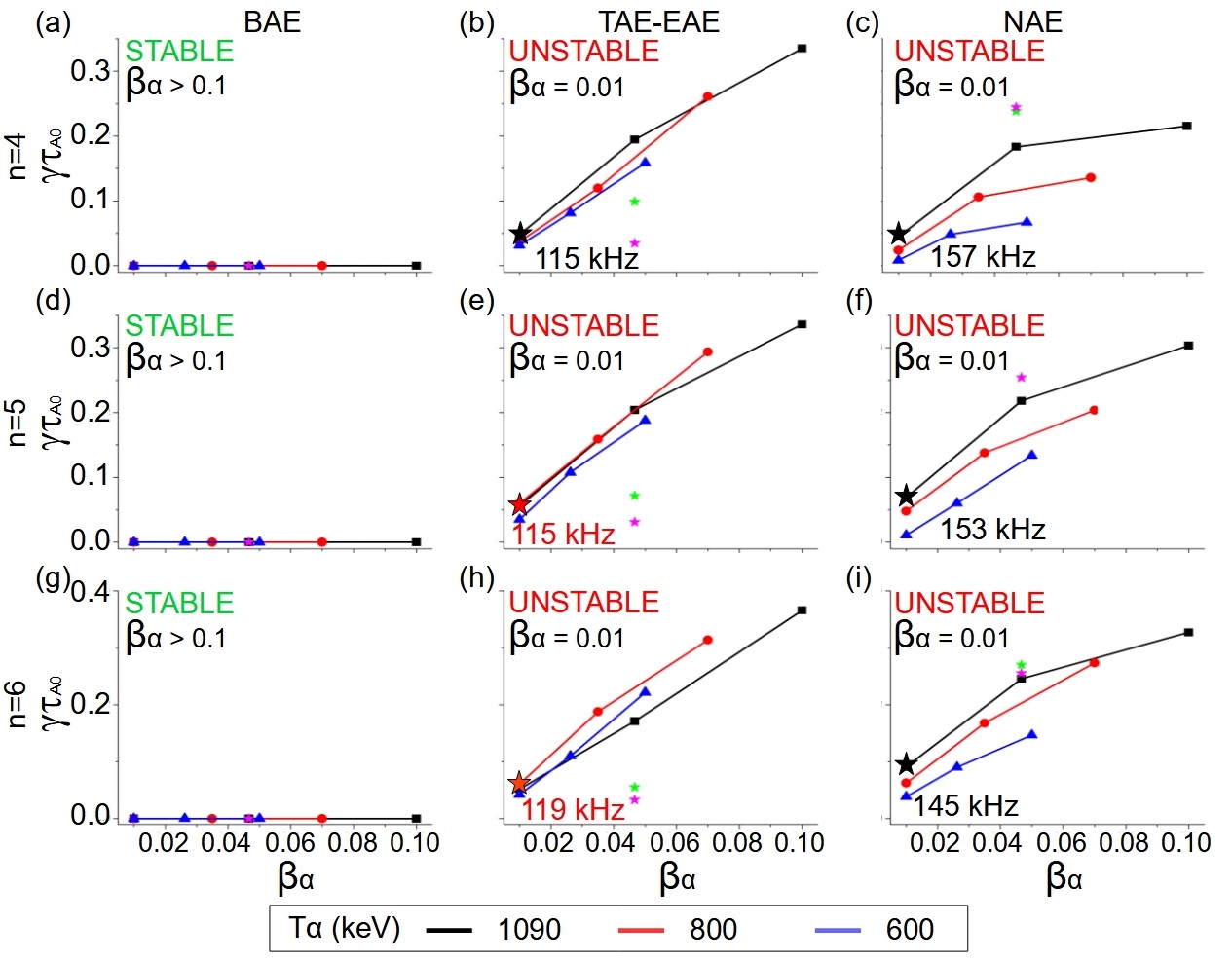}
\caption{Growth rate of $n=4$ to $6$ dominant modes for different EP energies and $\beta$ values destabilized by the $\alpha$ particles. The green/red capital letter in the top of the panel indicate if the mode is stable/unstable for the expected $\alpha$ particle population in the reference model. The colored stars indicate the modes  with the largest growth rate and lower $\beta_{\alpha}$ threshold) including the mode frequency. The $\beta_{\alpha}$ required to destabilize AE is also indicated. The $\beta_{\alpha}$ required to destabilize AE is also indicated. The green (pink) star indicates the growth rate of the AEs destabilized by $\alpha$ particles with $T_{\alpha}=2000$ ($3000$) keV and $\beta_{\alpha} = 0.04667$.}\label{FIG:6}
\end{figure}

Figures~\ref{FIG:5} and ~\ref{FIG:6} indicate the growth rate and frequency of the dominant $n=1$ to $6$ AEs destabilized by the $\alpha$ particles for different energies and $\beta$ values. The simulations indicate that $n=1$ to $3$ BAEs with $f = 52 - 61$ kHz are destabilized by type I and II $\alpha$ particles if the $\beta_{\alpha} \ge 0.01$. In addition, $n=1$ to $6$ TAE/EAEs ($f = 68 - 119$ kHz) and NAEs ($f > 125$ kHz) are unstable, triggered by type I and II $\alpha$ particles if the EP $\beta_{\alpha} \ge 0.01$. The simulations with $T_{\alpha}=2000$ and $3000$ keV show a decrease of the $n=1$ to $6$ BAEs and TAEs growth rate (except $n=1$ and $2$ TAEs) although an increase of the NAEs growth rate. Nevertheless, finite Larmor radius effects strongly damp high frequency AEs (please see section \ref{sec:Multiple} for more info), thus the general trend indicates a decrease of the AE growth rate as $T_{\alpha}$ increases above $1090$ keV. Consequently, the analysis is limited to $\alpha$ particles with $T_{\alpha} \le 1090$ keV.

\begin{figure}[h!]
\centering
\includegraphics[width=0.5\textwidth]{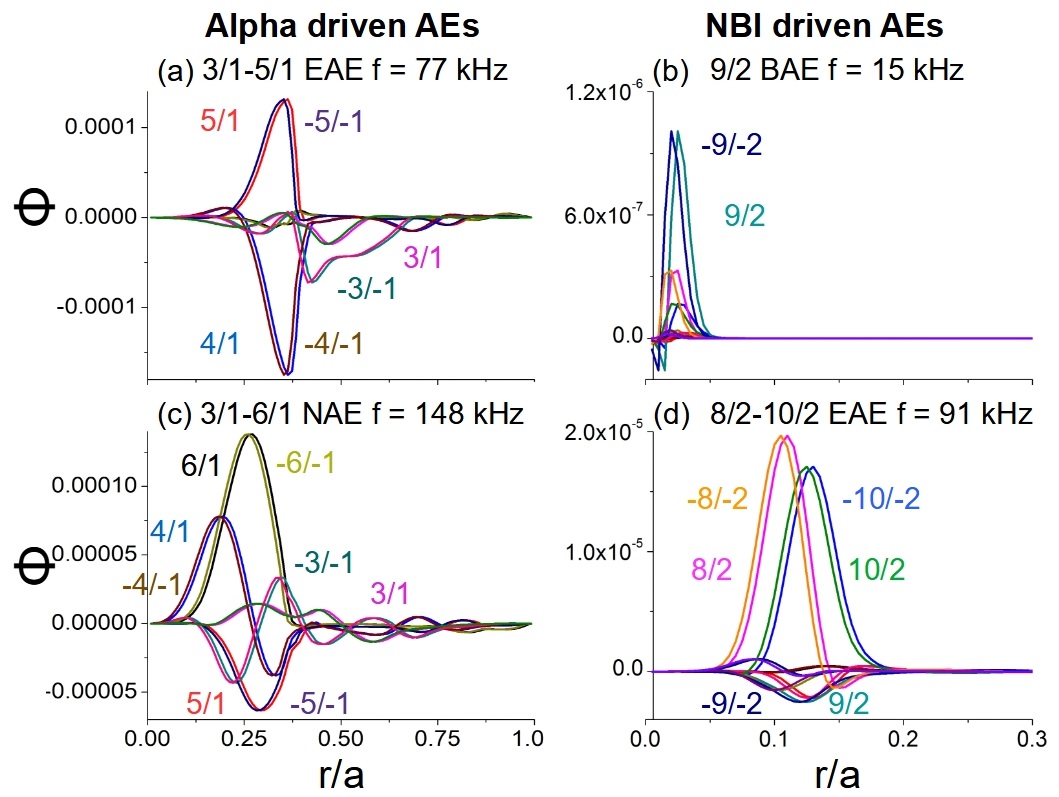}
\caption{Eigenfunction of AEs destabilized by $\alpha$ particles (a) $3/1-5/1$ EAE and (c) $3/1-6/1$ NAE. Eigenfunction of AEs destabilized by NBI EP (b) $9/2$ BAE and (d) $8/2-10/2$ EAE.}\label{FIG:7}
\end{figure}

Figure~\ref{FIG:7} shows some eigenfunction examples of dominant AEs destabilized by NBI EP and $\alpha$ particles. The AEs induced by the $\alpha$ particles are located in the inner-middle plasma region and the mode amplitude peaks at $r/a = 0.25 - 0.35$. The AEs are destabilized due to the gradient of the $\alpha$ particle density profile between $r/a = 0.2 - 0.5$, exceeding the stabilizing effect of the magnetic shear between $r/a = 0.35-0.5$. On the other hand, the AEs induced by the NBI EP are triggered near the magnetic axis because the NBI is injected on-axis, leading to a sharp slope of the NBI EP density profile between $r/a = 0.0 - 0.2$, combined with the weak stabilizing effect of the magnetic shear at $r/a < 0.35$ where the safety factor profile is almost flat.

In summary, the plasma of CFETR steady state operations with the ITB located at $r/a \approx 0.45$ is stable with respect to the EP driven by the NBI, except for marginally unstable $n=3$ and $n=5$ BAEs and $n=6$ EAE located near the magnetic axis. On the other hand, $n=1$ to $3$ BAEs, $n=1$ to $6$ TAE/EAEs and NAEs are destabilized by type I and II $\alpha$ particles in the inner-middle plasma region ($r/a = 0.25 - 0.35$). Consequently, there is an overlapping regarding the radial location and frequency range of the AEs induced by the resonant $\alpha$ particles along the thermalization process. Thus, the radial transport of the $\alpha$ particles could be enhanced in the inner-middle plasma region, leading to the outward flux of the $\alpha$ particles population before thermalization, impairing the performance of the device. It should be mentioned that other AEs with lower growth rates regarding the dominant AEs are also destabilized. These sub-dominant AEs lead to a smaller limitation of the device performance. This is the case of the reverse shear AEs (RSAE). The reason why the RSAE are sub-dominant modes is because the radial location of the largest EP density gradient and the reverse shear regions are not coincident in cases A and B, the EP density gradient is displaced inward regarding the q minima / maxima of the reverse magnetic shear region. Consequently, the free energy to destabilize the RSAE is smaller regarding other AEs triggered in the inner plasma region. More information about the sub-dominant modes is provided in sections \ref{sec:Multiple} and \ref{sec:High}. 

\subsection{Case B: ITB in the outer plasma region}

\begin{figure}[h!]
\centering
\includegraphics[width=0.5\textwidth]{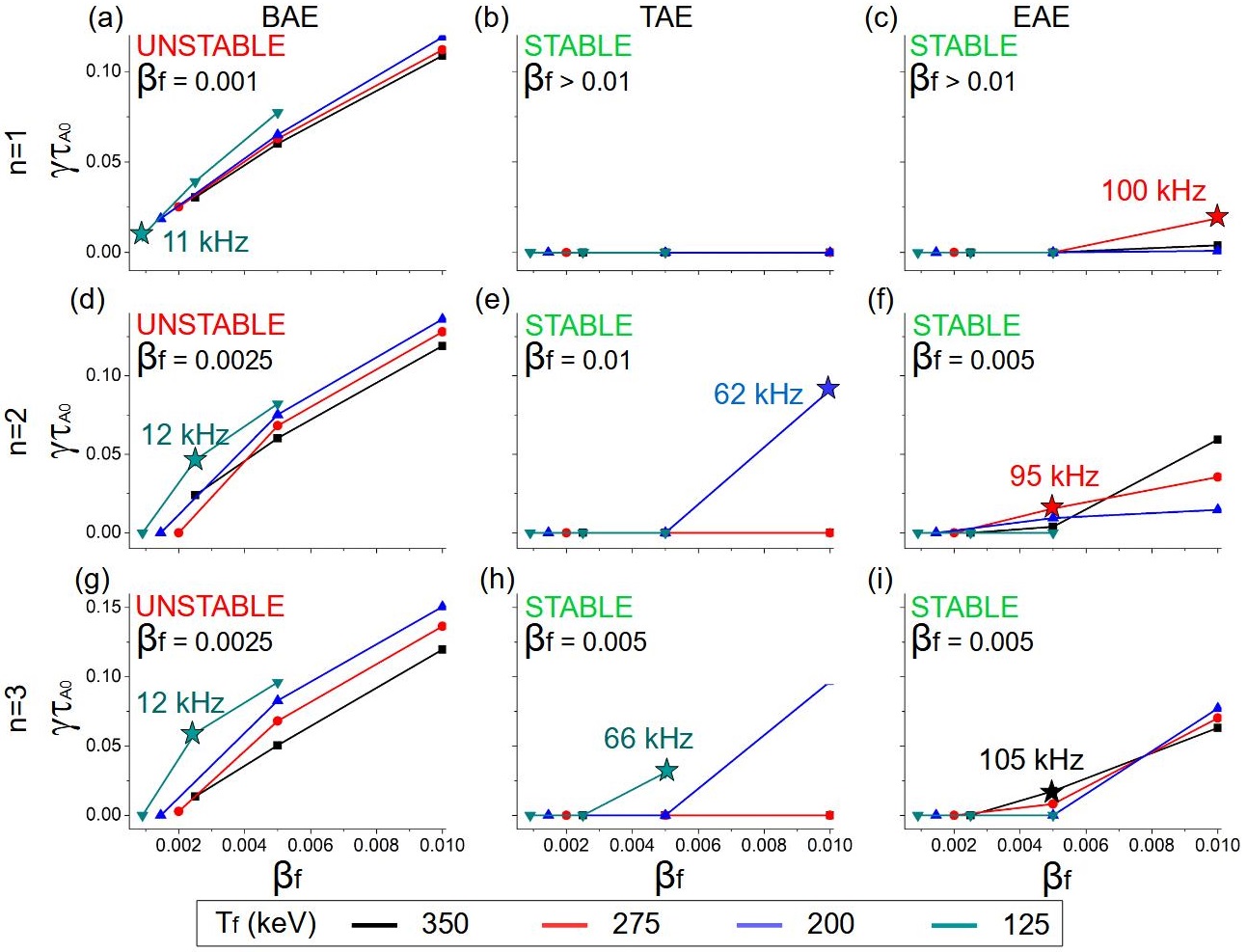}
\caption{Growth rate of $n=1$ to $3$ dominant modes for different EP energies and $\beta$ values destabilized by NBI driven EP. The green/red capital letter in the top of the panel indicate if the mode is stable/unstable for the reference operational regime of the NBI. The colored stars indicate the modes  with the largest growth rate and lower EP $\beta$ threshold, including the mode frequency. The NBI EP $\beta$ required to destabilize AE is also indicated.}\label{FIG:8}
\end{figure}

\begin{figure}[h!]
\centering
\includegraphics[width=0.5\textwidth]{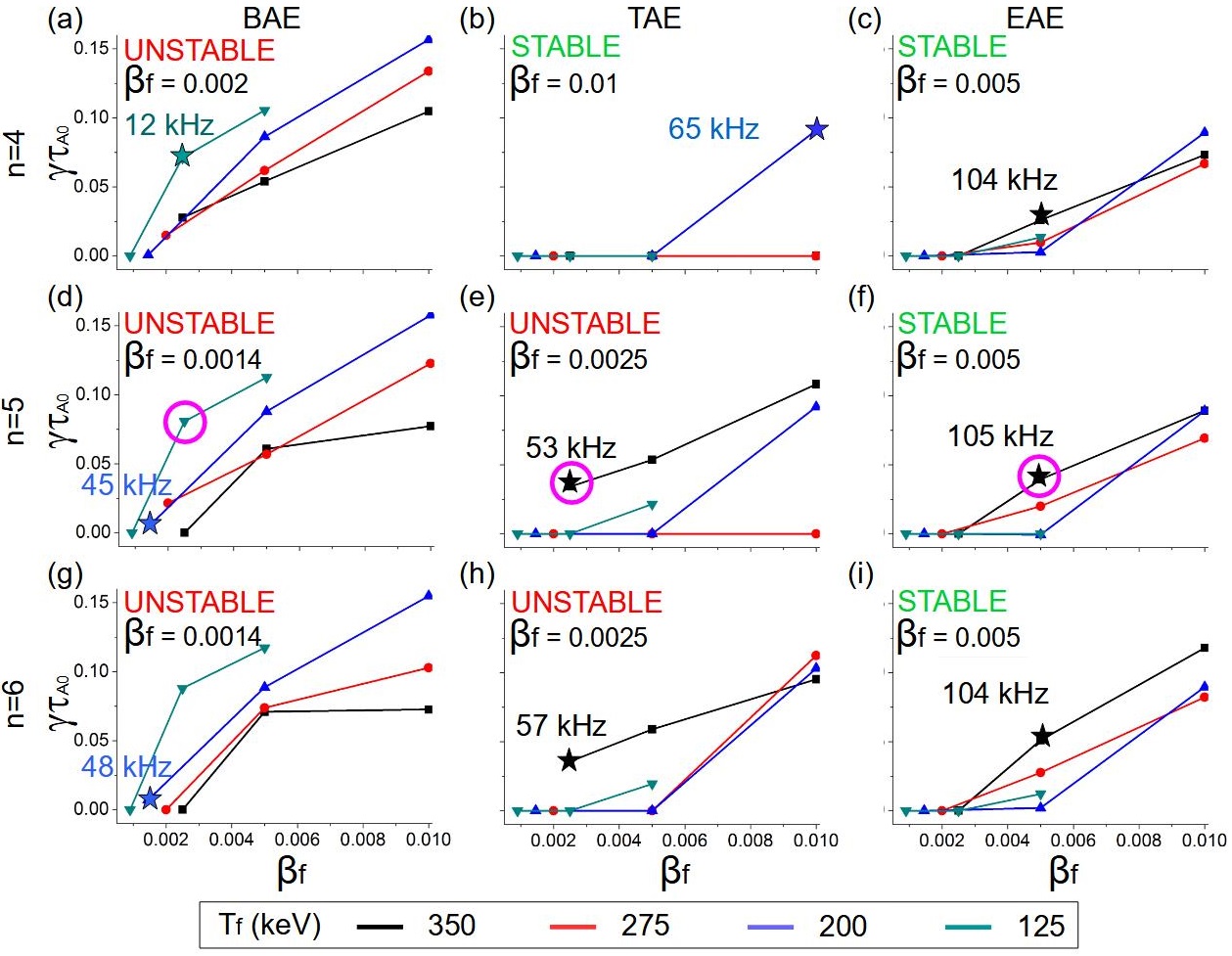}
\caption{Growth rate of $n=4$ to $6$ dominant modes for different EP energies and $\beta$ values destabilized by NBI driven EP. The green/red capital letter in the top of the panel indicate if the mode is stable/unstable for the reference operational regime of the NBI. The colored stars indicate the modes with the largest growth rate and lower EP $\beta$ threshold, including the mode frequency. The pink circles indicate the modes whose eigenfunction is plotted in fig. 11. The NBI EP $\beta$ required to destabilize AE is also indicated.}\label{FIG:9}
\end{figure}

Figures~\ref{FIG:8} and~\ref{FIG:9} show the growth rate and frequency of the $n=1$ to $6$ dominant modes destabilized by NBI EP with different energies and $\beta$ values. The simulations indicate the destabilization of $n=1$ to $4$ BAEs with $f=14$ kHz and $n=5$ to $6$ BAEs with $f \approx 45$ kHz by type II NBI EPs as well as $n=5$ and $6$ TAEs with $f \approx 55$ kHz by type I NBI EPs. It should be recalled that the NBI EP $\beta$ in the case B is $0.00257$, almost two times larger with respect to case A, large enough to exceed the stability limit of the AEs. In addition, if the NBI EP $\beta$ is increased above the reference model, $n=3$ TAE with $f=66$ kHz as well as $n=2$ to $6$ EAEs with $f = 95 - 105$ kHz are unstable for $\beta_{f} = 0.005$, mainly triggered by type I EPs.

\begin{figure}[h!]
\centering
\includegraphics[width=0.5\textwidth]{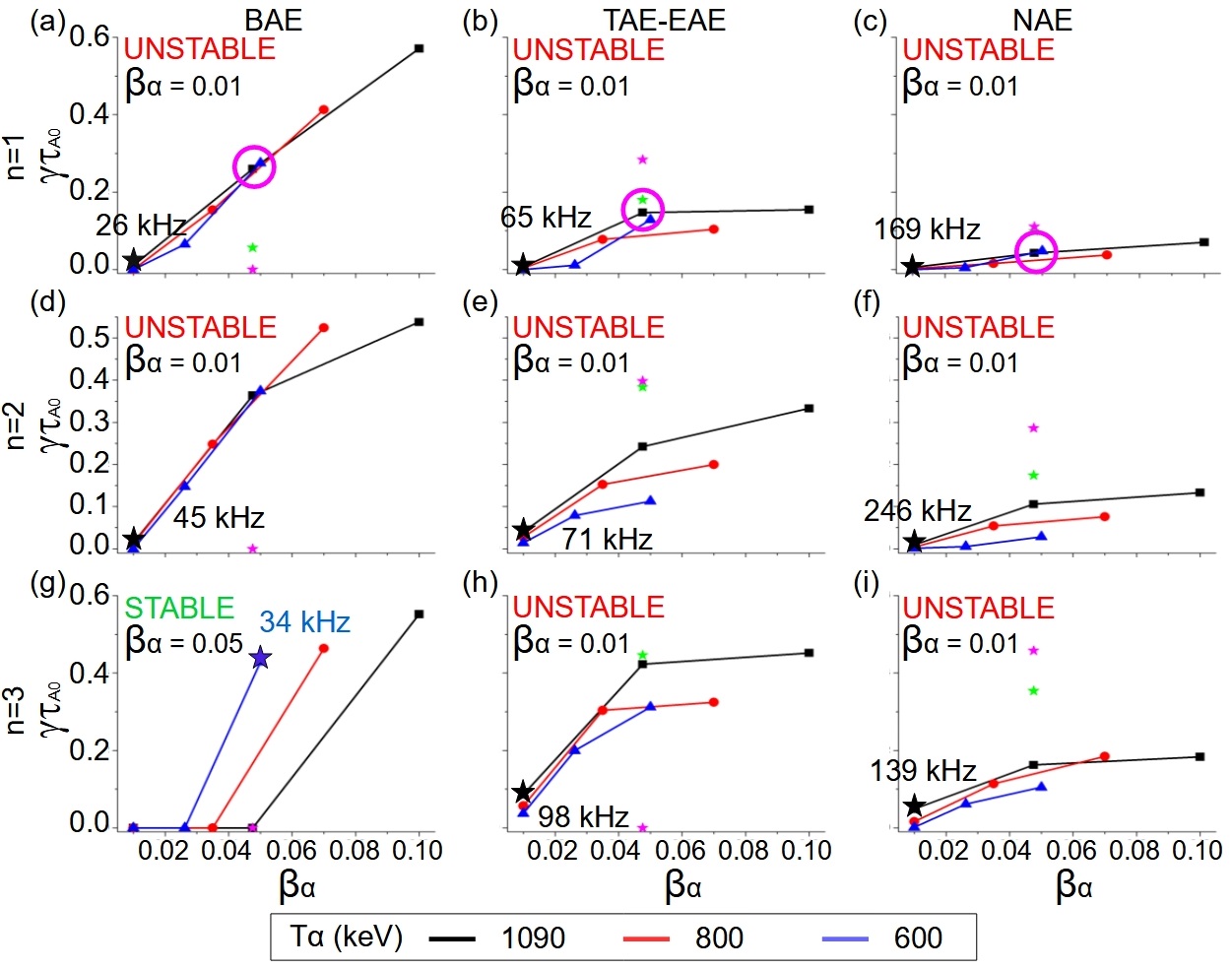}
\caption{Growth rate of $n=1$ to $3$ dominant modes for different EP energies and $\beta$ values destabilized by the $\alpha$ particles. The green/red capital letter in the top of the panel indicate if the mode is stable/unstable for the expected $\alpha$ particle population in the reference model. The colored stars indicate the modes with the largest growth rate and lower $\beta_{\alpha}$ threshold, including the mode frequency. The pink circles indicate the modes whose eigenfunction is plotted in fig. 11. The $\beta_{\alpha}$ required to destabilize AEs is also indicated. The green (pink) star indicates the growth rate of the AEs destabilized by $\alpha$ particles with $T_{\alpha}=2000$ ($3000$) keV and $\beta_{\alpha} = 0.04667$.}\label{FIG:10}
\end{figure}

\begin{figure}[h!]
\centering
\includegraphics[width=0.5\textwidth]{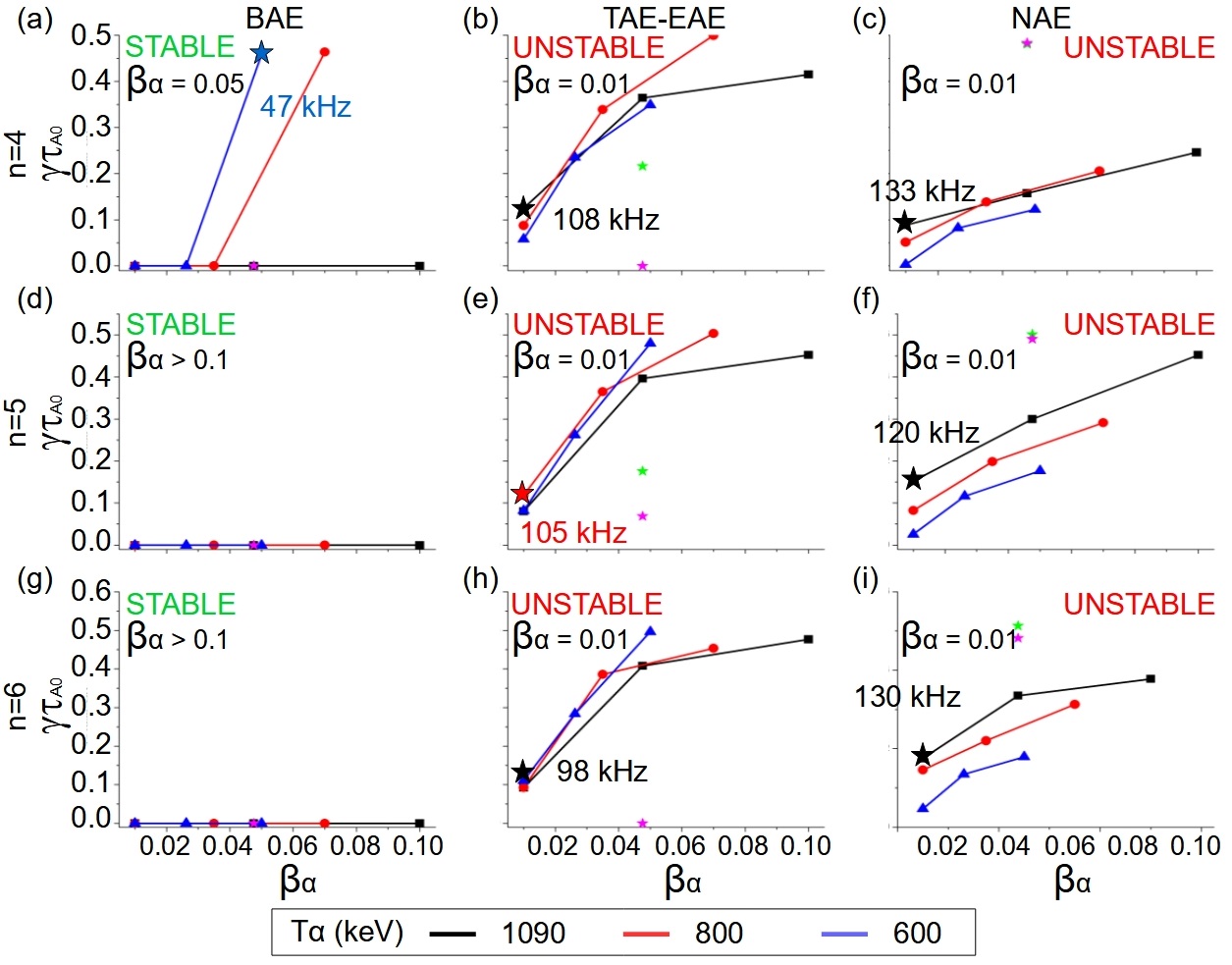}
\caption{Growth rate of $n=4$ to $6$ dominant modes for different EP energies and $\beta$ values destabilized by the $\alpha$ particles. The green/red capital letter in the top of the panel indicate if the mode is stable/unstable for the expected $\alpha$ particle population in the reference model. The colored stars indicate the modes  with largest growth rate and lower $\beta_{\alpha}$ threshold, including the mode frequency. The pink circles indicate the modes whose eigenfunction is plotted in fig. 11. The $\beta_{\alpha}$ required to destabilize AE is also indicated. The green (pink) star indicates the growth rate of the AEs destabilized by $\alpha$ particles with $T_{\alpha}=2000$ ($3000$) keV and $\beta_{\alpha} = 0.04667$.}\label{FIG:11}
\end{figure}

Figures~\ref{FIG:10} and~\ref{FIG:11} show the growth rate and frequency of the $n=1$ to $6$ dominant modes for different energies and $\beta$ value destabilized by the $\alpha$ particles. The $n=1$ to $2$ BAEs with $f = 26 - 45$ kHz and $n=1$ to $6$ TAE/EAEs with $f = 65 - 108$ kHz are destabilized by type I and II $\alpha$ particles as well as $n=1$ to $6$ NAEs with $f > 120$ kHz mainly triggered by type I $\alpha$ particles. The AEs growth rate is higher in case B with respect to case A. The reason why the case B shows a worse AE stability is because the strongest gradient of the $alpha$ particle density profile is located between $r/a = 0.5 - 0.6$, a plasma region where the q profile has a turning point and the stabilizing effect of the magnetic shear is smaller. The simulations with $T_{\alpha} > 1090$ keV also shows a lower AE growth rate regarding the simulations with $T_{\alpha} \le 1090$.

\begin{figure}[h!]
\centering
\includegraphics[width=0.5\textwidth]{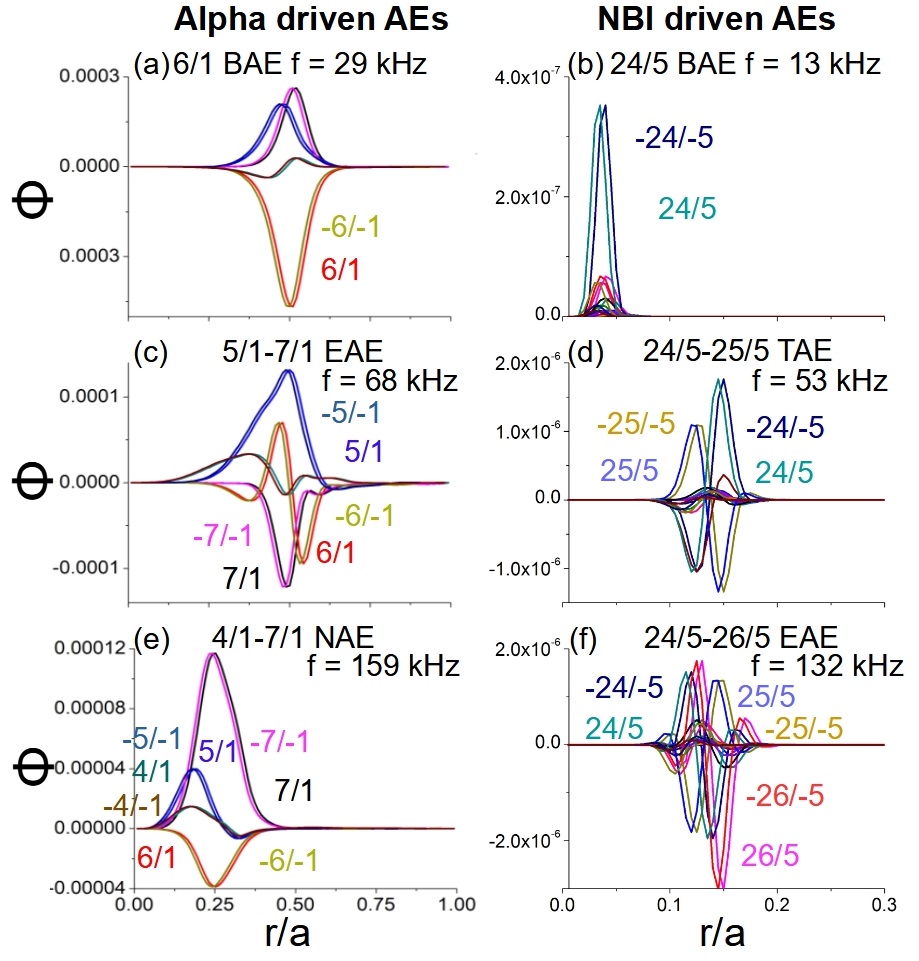}
\caption{Eigenfunction of AEs destabilized by $\alpha$ particles (a) $6/1$ BAE, (c) $5/1-7/1$ EAE and (e) $4/1-7/1$ NAE. Eigenfunction of AEs destabilized by NBI EP (b) $24/5$ BAE, (d) $24/5-25/5$ TAE and (f) $24/5-26/5$ EAE.}\label{FIG:12}
\end{figure}

Figure~\ref{FIG:12} shows some eigenfunction examples of dominant AE destabilized by NBI EP and $alpha$ particles. The AEs triggered by the $\alpha$ particles are located in the inner-middle plasma region. On the other, the BAEs destabilized by the NBI EP are located near the magnetic axis. Consequently, $\alpha$ particles and NBI EP populations resonate at closer radial location, thus the resulting AE stability of the plasma can be affected by multiple EP population effects that will be evaluated in the following section \cite{57,58}.

In summary, the plasma of CFETR steady state operations with the ITB located at $r/a \approx 0.6$ is unstable to $n=1$ to $6$ BAE by type II NBI EPs and $n=5$ and $6$ TAE by type I NBI EPs. In addition, $n=1$ to $2$ BAEs as well as $n=1$ to $5$ TAE/EAEs are destabilized by type I and II $\alpha$ particles, as well as $n=1$ to $n=6$ NAEs by type I $\alpha$ particles. Comparing the AE stability of cases A and B, the AEs growth rate is larger in case B because the stabilizing effect of the magnetic shear in the middle plasma region, where the AEs are triggered, is weaker.

\section{Simulations with multiple EP population for low $n$ AEs \label{sec:Multiple}}

This section is dedicated to study the AE stability if both EP populations, EP NBI and $\alpha$ particles, are included in the simulations, hence multiple EP population effect on the AE stability can be identified. The study is performed for type I $\alpha$ particles and NBI EP ($T_{\alpha} = 1090$ keV and $T_{f} = 350$ keV) as well as for type II $\alpha$ particles and NBI EP ($T_{\alpha} = 600$ keV and $T_{f} = 200$ keV). In addition, the damping effect caused by the EP and thermal ion Finite Larmor Radius (FLR) are included in the simulations. The Larmor radius used in the study is $0.025$ m for the $\alpha$ particles, $0.01$ m for the NBI EP and $0.003$ m for the thermal ions. It should be mentioned that the Larmor radius depends on the magnetic field intensity and perpendicular energy of the EP / thermal ions. For simplicity, the model approximates the Larmor radius at the plasma region where the AEs are destabilized (inner-middle plasma region). The aim of the present study is providing an estimation of the FLR effect on the AE growth rate, although further analysis using more sophisticated approximations are required to improve the prediction accuracy. There is more info about the implementation of the EP FLR effects in the appendix of \cite{59}.

\subsection{Type I EPs}

\begin{figure}[h!]
\centering
\includegraphics[width=0.5\textwidth]{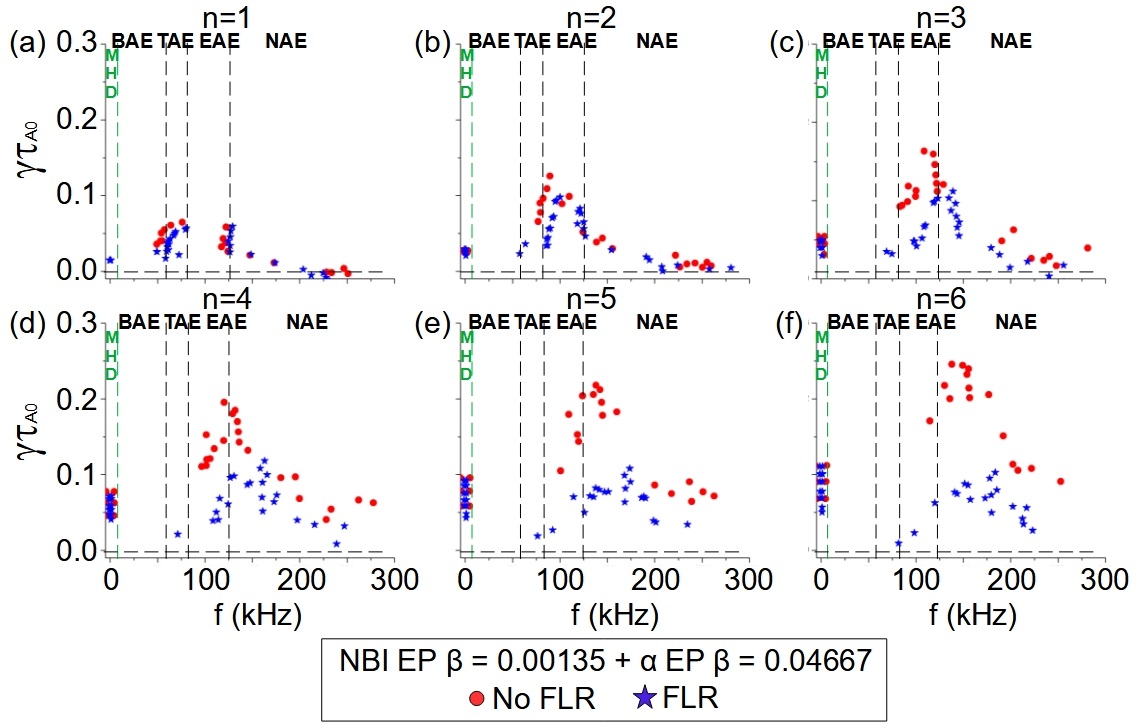}
\caption{Normalized growth rate and frequency of the dominant and sub-dominant modes for the case A and type I $\alpha$ particles + NBI EP. The vertical green dashed line indicates the range of frequencies of the pressure gradient driven modes (label MHD, low frequency modes). The horizontal dashed black line separates the stable modes (negative growth rate) and unstable modes. The red dots indicates the simulations without FLR damping effects, green stars with EP FLR and blue triangles with thermal ion FLR effects. The dashed black vertical lines indicate the frequency range of the different AE family gaps between the magnetic axis and the middle plasma.} \label{FIG:13}
\end{figure}

\begin{figure}[h!]
\centering
\includegraphics[width=0.5\textwidth]{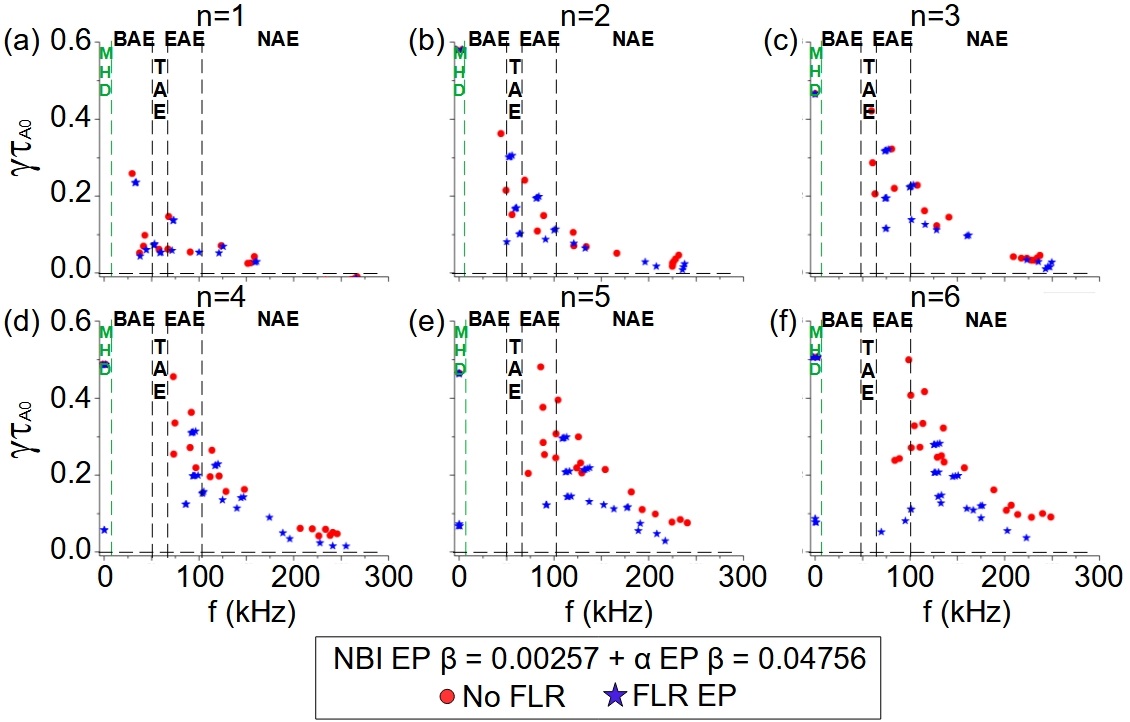}
\caption{Normalized growth rate and frequency of the dominant and sub-dominant modes for the case B and type I $\alpha$ particles + NBI EP. The vertical green dashed line indicates the range of frequencies of the pressure gradient driven modes (label MHD, low frequency modes). The horizontal dashed black line separates the stable modes (negative growth rate) and unstable modes. The red dots indicates the simulations without FLR damping effects, green stars with EP FLR and blue triangles with thermal ion FLR effects. The dashed black vertical lines indicate the frequency range of the different AE family gaps between the magnetic axis and the middle plasma.} \label{FIG:14}
\end{figure}

Figures~\ref{FIG:13} and~\ref{FIG:14} show the growth rate and frequency of the dominant and sub-dominant modes for the cases A and B, respectively. The growth rate of the AEs destabilized in the case B is larger with respect to the AEs triggered in the case A. In addition, the dominant AEs destabilized in the case A are in the frequency range of the EAE ($n=1$ to $2$) and NAE ($n=3$ to $6$) gaps, although in the case B the dominant modes are $n=1$ BAE, $n=2$ TAE, $n=3$ to $4$ EAEs and $n=5$ to $6$ NAEs. Consequently, the FLR damping effect largely reduce the growth rate of the dominant modes in the case A with respect to the case B, as can be observed comparing the AEs calculated in the simulation without FLR effects (red dots) with the AE in the simulations with FLR effects (blue stars). It should be recalled that the FLR damping effects are enhanced if the mode width is closer to the EP or thermal ions Larmor radius, leading to a stronger orbit averaging effect of the EP and thermal ion gyro-motion acting on the electromagnetic fields of the instability; this is the case of high frequency AE as EAEs and NAEs, showing radially localized eigenfuctions regarding the BAEs or TAEs eigenfunctions. Thus, if the simulation includes the FLR effect, the growth rate of the dominant modes in the case A decreases by $50 - 70 \%$ (see $n=3$ to $n=6$ NAEs), although the decrease of the growth rate for the dominant $n=1$ TAE as well as the $n=2$ EAE is smaller than $10 - 25 \%$. The thermal ion FLR effect leads to a smaller decrease of the modes growth rate with respect to the EP FLR effects. On the other hand, the thermal ion FLR effect is dominant for the modes with the largest frequencies, above $200$ kHz. The effect of multiple EP populations in the case A is negligible for type I $\alpha$ particles because the growth rate and frequency of the dominant and sub-dominant modes is almost the same as compared to the simulation with only $\alpha$ particles. On the other hand, case B shows a slight increase of the $n=4$ to $6$ TAE/EAE and NAE growth rate, around $10-25 \%$.

\subsection{Type II EPs}

\begin{figure}[h!]
\centering
\includegraphics[width=0.5\textwidth]{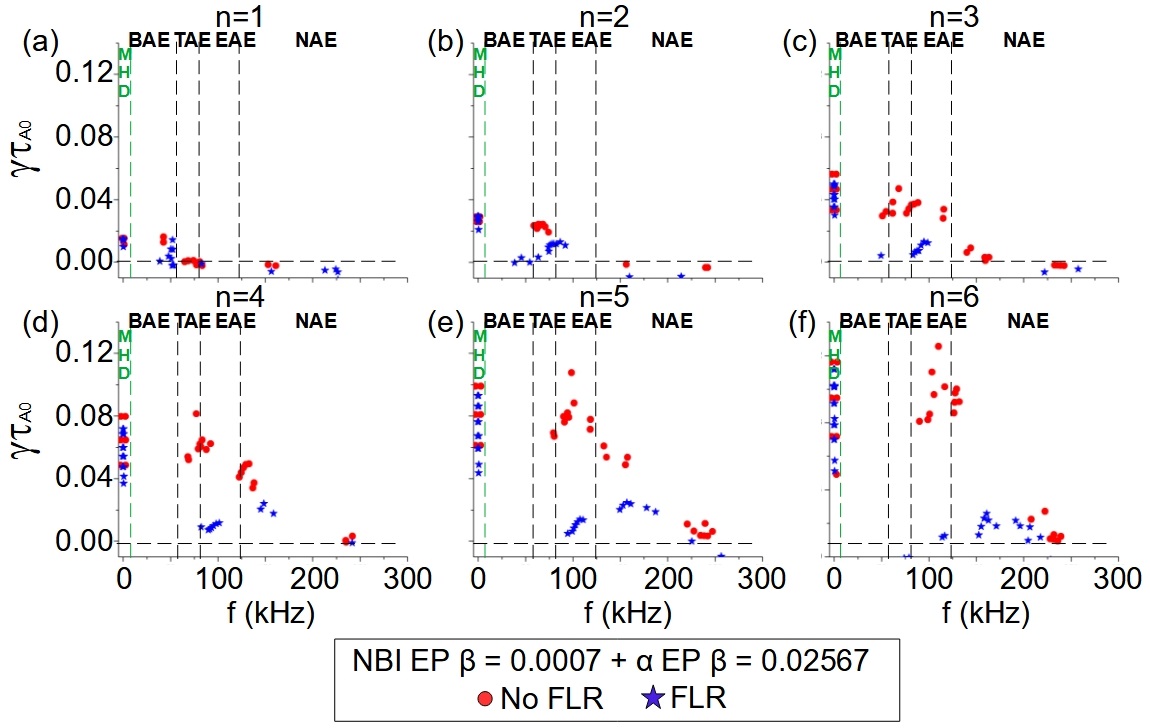}
\caption{Normalized growth rate and frequency of the dominant and sub-dominant modes for the case A and type II $\alpha$ particles + NBI EP. The vertical green dashed line indicates the range of frequencies of the pressure gradient driven modes (label MHD, low frequency modes). The horizontal dashed black line separates the stable modes (negative growth rate) and unstable modes. The red dots indicates the simulations without FLR damping effects, green stars with EP FLR and blue triangles with thermal ion FLR effects. The dashed black vertical lines indicate the frequency range of the different AE family gaps between the magnetic axis and the middle plasma.} \label{FIG:15}
\end{figure}

\begin{figure}[h!]
\centering
\includegraphics[width=0.5\textwidth]{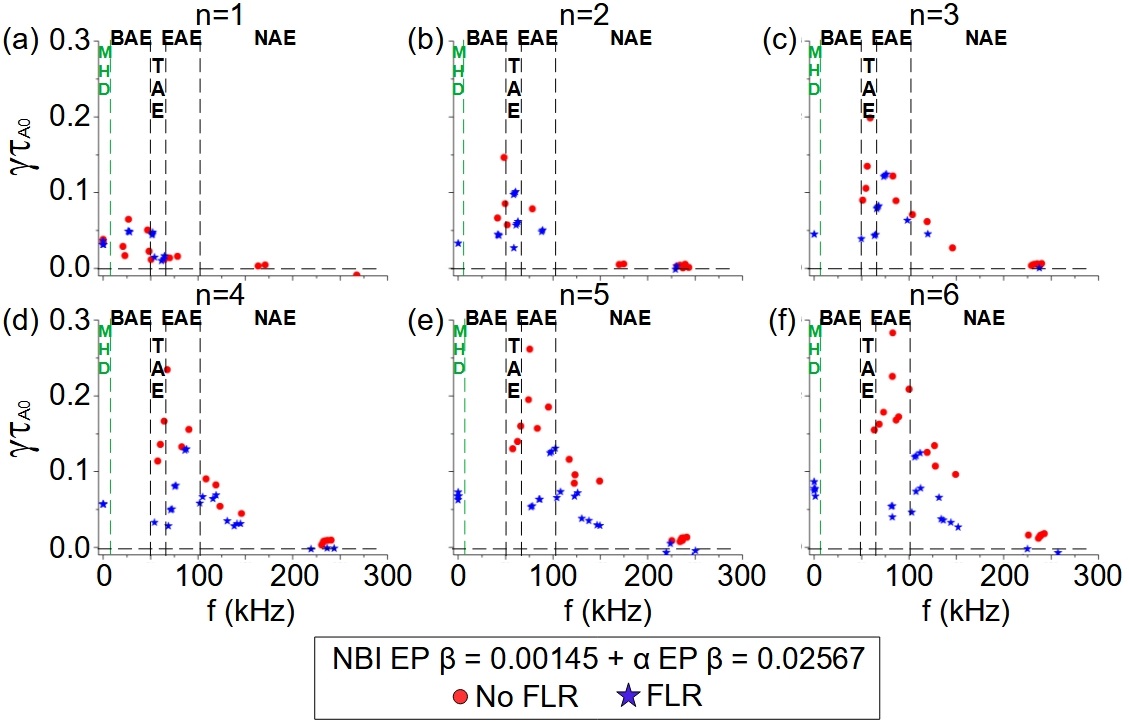}
\caption{Normalized growth rate and frequency of the dominant and sub-dominant modes for the case B and type II $\alpha$ particles + NBI EP. The vertical green dashed line indicates the range of frequencies of the pressure gradient driven modes (label MHD, low frequency modes). The horizontal dashed black line separates the stable modes (negative growth rate) and unstable modes. The red dots indicates the simulations without FLR damping effects, green stars with EP FLR and blue triangles with thermal ion FLR effects. The dashed black vertical lines indicate the frequency range of the different AE family gaps between the magnetic axis and the middle plasma.} \label{FIG:16}
\end{figure}

Figures~\ref{FIG:15} and~\ref{FIG:16} show the growth rate and frequency of the dominant and sub-dominant modes for the cases A and B, respectively. Including the FLR effect in case A simulations leads to a decrease of the $n=1$ BAE growth rate around $5 \%$, $n=2$ and $3$ TAE/EAEs growth rate by $50 - 60 \%$ and the $n=4$ to $6$ NAE by $60 - 85 \%$. Again, the effect of the thermal ion FLR damping is smaller regarding the EP FLR. Multiple EP population effects are weak because the growth rate and frequency of the AEs in the simulations with $\alpha$ particles + NBI EP and the simulations with only $\alpha$ particles are almost the same. Likewise, the effect of multiple EP populations is weak on case B except for the $n=1$ BAE that shows a decrease of $15 \%$ in the growth rate. Consequently, the NBI EP yields a stabilizing effect on the AEs induced by the $\alpha$ particles during the slowing down process, particularly on low $n$ and low frequency modes.

\section{Analysis of high n mode stability \label{sec:High}}

\begin{figure}[h!]
\centering
\includegraphics[width=0.5\textwidth]{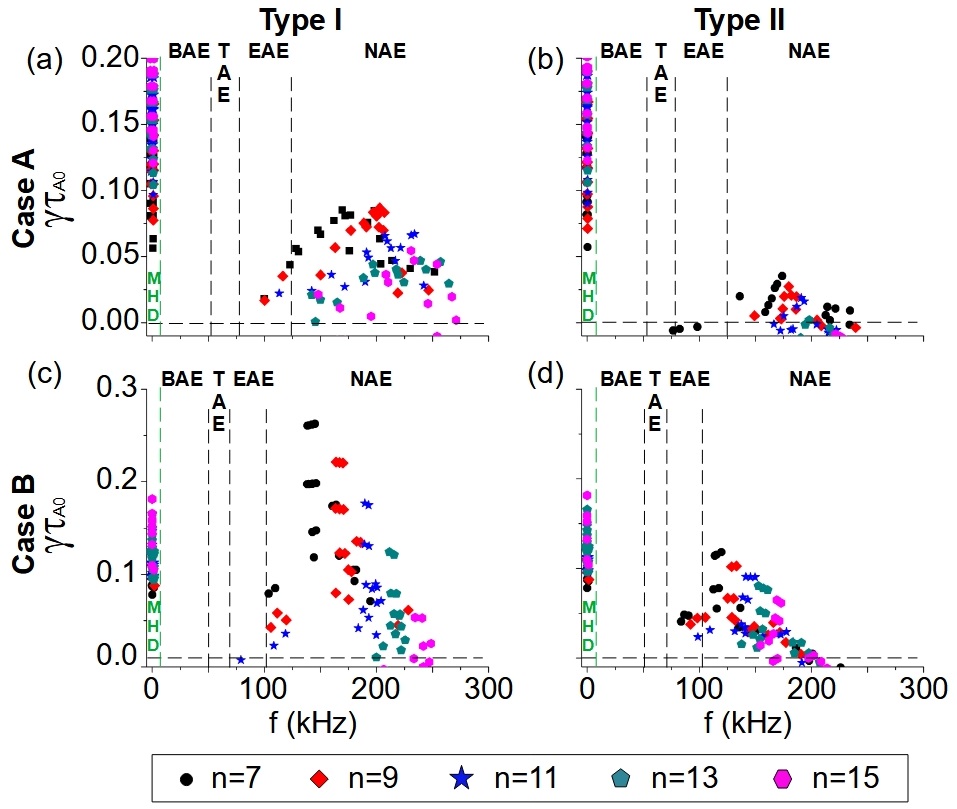}
\caption{Growth rate and frequency of the $n=7$ (black circle), $9$ (red diamond), $11$ (blue star), $13$ (dark cyan pentagon) and $15$ (pink hexagon) toroidal families. Case A (a) type I and (b) type II EPs. Case B (c) type I and (d) type II EPs. The horizontal dashed black line separates the stable modes (negative growth rate) and unstable modes. The dashed black vertical lines indicate the frequency range of the different AE family gaps between the magnetic axis and the middle plasma.} \label{FIG:17}
\end{figure}

This section is dedicated to study the stability of high $n$ AE ($n > 6$). The simulations include multiple EP populations and FLR damping effects. Figure~\ref{FIG:17} shows the growth rate and frequency of the dominant and subdominant modes of $n=7$, $9$, $11$, $13$ and $15$ toroidal families. 

The simulations indicate a decrease of the growth rate and an increase of the frequency of the dominant mode as the toroidal mode number increases. All the dominant modes of the high $n$ modes analyzed are NAE. The $n=14$ to $15$ NAEs are stable in the simulations of case A for type II EPs.

\begin{figure}[h!]
\centering
\includegraphics[width=0.5\textwidth]{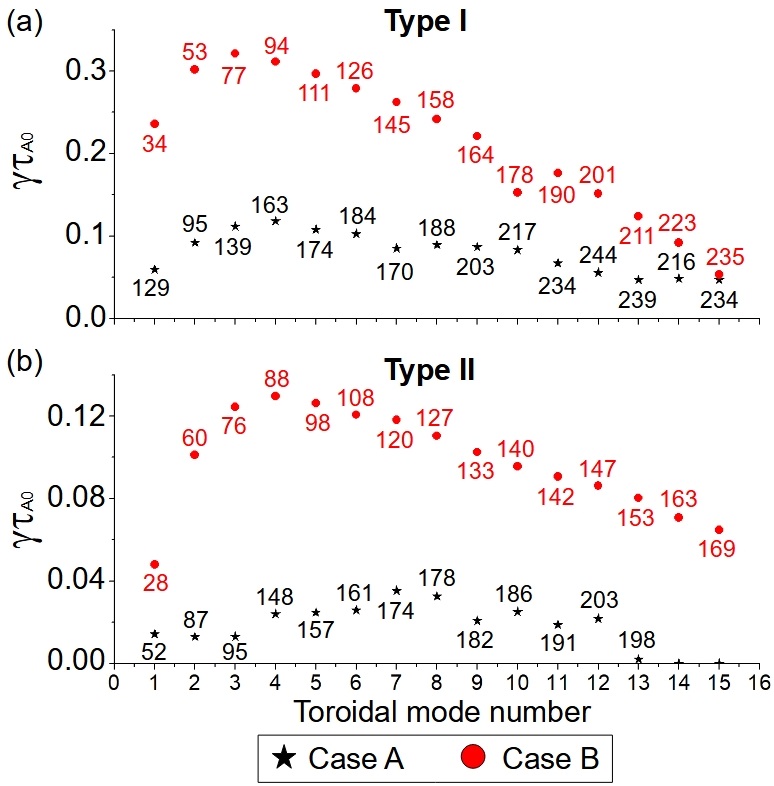}
\caption{Growth rate of the $n=1 - 15$ toroidal families dominant mode for the case A (black stars) and Case B (red circles). (a) Type I $\alpha$ + NBI EP (b) type II $\alpha$ particles + NBI EP. The frequency of the mode is indicated.} \label{FIG:18}
\end{figure}

Figure~\ref{FIG:18} shows the growth rate of the dominant mode in the simulations including multiple EP populations and FLR effects for the cases A and B as well as for type I and II EPs. The modes with the largest growth rate are low $n$ modes, $n=4$ NAE with $f = 163$ kHz in the case A for type I EPs, $n=7$ NAE with $f=174$ kHz in the case A for type II EPs, $n=3$ TAE/EAE with $f=98$ kHz in the case B for type I EPs as well as $n=4$ TAE/EAE with $f=88$ kHz in the case B for type II EPs. The dominant mode growth rate decreases and the frequency increases with the toroidal mode number for the high $n$ modes. The decrease of the high $n$ modes growth rate relative to the low $n$ modes can be explained by the strong damping caused by FLR effects, because the high $n$ modes show an narrower eigenfunction relative to the low $n$ modes, particularly for the high frequency modes in the range of the NAEs.

Summarizing, the destabilization of low $n$ AEs in the CFETR configurations analyzed could degrade plasma heating efficiency of the tangential NBIs and $\alpha$ particle thermalization, due to unstable AEs in the frequency range of $28 - 184$ kHz. The role of the high $n$ AEs in the plasma heating efficiency is smaller than the low $n$ modes, showing a lower growth rate and destabilizing modes only in the frequency range of the NAEs ($f > 127$ kHz), which are strongly damped by the EP FLR effects. It should be noted that the growth rate of high n modes are further reduced if the thermal ion FLR effects are also included in the simulations, leading to the NAE stabilization for a lower n mode number.

\section{Conclusions and discussion \label{sec:conclusions}}

The stability of low $n$ modes in CFETR plasma for steady state operations is analyzed using the gyro-fluid code FAR3d. Two configurations are explored with respect to the radial location of the internal transport barrier. 

The simulations for a single EP population indicate that the growth rate and $\beta$ threshold of the AEs destabilized by the NBI EP and $\alpha$ particles are smaller if the ITB is located at $r/a \approx 0.45$ than for the configuration with the ITB at $r/a \approx 0.6$. The plasma stability is improved in the configuration with the ITB located inward because the AEs are destabilized in a region with large magnetic shear. Consequently, the AEs eigenfunction is narrow and the free energy required to trigger the mode is higher.

The analysis of the continuum gaps between the inner-middle plasma region for both configurations shows a TAE gap with narrow frequency ranges, although wide EAEs that extend from $\approx 60$ to $\approx 125$ kHz. In addition, the upper frequency of the BAE gap is $\approx 50$ kHz and above $125$ kHz there are several NAE gaps. Consequently, the unstable AEs mainly belong to the Elliptic and $\beta$ induced AE families.

The simulations for the configuration with inward ITB show that the NBI EP can trigger marginally unstable $n=3$ and $n=5$ BAEs and an $n=6$ EAE located near the magnetic axis. In addition, $n=1$ to $3$ BAEs and $n=1$ to $6$ TAE/EAEs and NAEs can be destabilized by type I and II $\alpha$ particles; the modes are located at the inner-middle plasma region. It should be noted that the simulations indicate an excess of the AEs $\beta_{\alpha}$ threshold, thus the resonances induced by type I and II $\alpha$ particles can lead to an enhancement of the $\alpha$ particles transport, an outward flux of $\alpha$ particles before the full thermalization and a lower heating efficiency of the plasma core. 

The simulations for the configuration with an outward ITB indicate the destabilization of $n=1$ to $6$ BAE near the magnetic axis by type II NBI EP and $n=5$ and $6$ TAE by type I NBI EP at the middle plasma region. Type I and II $\alpha$ particles trigger $n=1$ to $2$ BAEs and $n=1$ to $6$ TAE/EAEs as well as $n=1$ to $n=6$ NAEs by type II $\alpha$ particles; the modes are destabilized at the middle plasma region. Consequently, the AEs triggered in the middle of the plasma can lead to the enhancement of the $\alpha$ particles transport.

The simulations including multiple EP populations, $\alpha$ particles and NBI EP, show a negligible effect on the stability of the AEs triggered by $\alpha$ particles, that is to say, the growth rate and frequency of the AEs is almost the same with respect to simulations including only the destabilizing effect of the $\alpha$ particles. On the other hand, multiple EP simulations for type II EPs in the configuration with outward ITB indicate a stabilizing effect of the NBI EP on the $n=1$ BAE triggered by the $\alpha$ particles, leading to a growth rate $15 \%$ lower comparing the simulation with only $\alpha$ particles and the simulations with multiple EP populations. Similar multiple EP population effects were observed in TFTR plasma; AEs triggered by $\alpha$ particles were only destabilized after the beam injection was stopped \cite{60,61,62}. Recent analysis dedicated to the ITER device also predict multiple EP population effects in plasma with large populations of $\alpha$ particle and NBI EP \cite{57}. It should be noted that the multiple EP effect is different regarding the density profiles, $\beta$ and temperature of each EP population as it was discussed in previous studies \cite{58}.

Finite Larmor Radius effects on the EPs and thermal ions are also included in the simulations, leading to a decrease of the AEs growth rate up to $80 \%$. The decrease of the AEs growth rate is larger in the configuration with inward ITB because the dominant modes belong to high frequency families as EAEs and NAEs, although for the configuration with outward ITB only $n>3$ AEs are EAE/NAEs. That happens because the FLR damping effects are stronger on modes whose eigenfunction is narrow, particularly EAE/NAEs.

The analysis of high $n$ modes ($n=7$ to $15$) indicates that the modes leading to the strongest limitation of the plasma heating efficiency regarding the tangential NBI and $\alpha$ particle thermalization in the CFETR configuration analyzed are the low $n$ modes. The simulations show that the growth rate of the high $n$ AEs are largely reduced by the FLR effects, destabilized in the frequency range of the NAEs. Previous studies concluded that only high $n$ AEs are unstable in ITER, FIRE and IGNITOR devices \cite{63}, although the numerical model applied did not include FLR damping effects on NBI EP and alpha particles (only FLR effects on thermal ions); these are particularly strong for high frequency AEs with a narrow eigenfunction radial width. Regarding low / medium $n$ AEs, the stability of these modes depends on the operation scenario, showing smaller FLR damping effects compared to high $n$ modes if the eigenfunction width is large enough and the frequency is not very high (below $100$ kHz).

The present study indicates that the CFETR steady state operations exceed the stability limit of low $n$ AEs triggered by $\alpha$ particles, particularly BAEs and EAEs at the inner-middle plasma region. The formation of a reverse magnetic shear region if the ITB is located inward improves the plasma stability, reducing the growth rate of the AEs induced by the $\alpha$ particles and relaxing the $\beta$ threshold, because the AEs are triggered in a plasma region with strong magnetic shear. On top of that, the AEs are triggered in the inner plasma if the ITB is at $r/a \approx 0.45$, separated with respect to the loss cone region located at $r/a \approx 0.7$ in CFETR plasma \cite{64}. Also, the destabilizing effect of the NBI EP is small, showing a stabilizing effect on the AEs triggered by $\alpha$ particles during the slowing down process. The large outward fluxes of non thermalized $\alpha$ particles may be avoided to improve the heating efficiency of the device, thus the $\alpha$ particles transport could be minimized, particularly if the transport is enhanced by the overlapping of different resonances. Consequently, the AEs $\beta$ threshold must be relaxed and the growth rate reduced due to the stabilizing effect provided by the magnetic shear, testing configurations with a large gradient of the safety factor profile in the plasma regions where the AEs are triggered. In addition, the stabilizing effect of the NBI EP on the AEs induced by the $\alpha$ particles should be analyzed, exploring different NBI operational regimes regarding the NBI deposition region, injection intensity and energy, identifying configurations that induce damping effects by multiple EP populations.

\ack
This work is supported by the project $2019-T1/AMB-13648$ founded by the Comunidad de Madrid, the National Natural Science Foundation of China under Grant No. $11975276$, Anhui Provincial Natural Science Foundation No. $2008085J04$, National Key Research and Development Program of China No. $2019YFE03020004$, Anhui Provincial Key R$\&$D Programmes No. $202104b11020003$ and the Excellence Program of Hefei Science Center CAS No. 2021HSC-UE015. This work has been also supported by Comunidad de Madrid (Spain) - multiannual agreement with UC3M (“Excelencia para el Profesorado Universitario” - EPUC3M14 ) - Fifth regional research plan 2016-2020. The authors also acknowledge J.M. Reynolds and V. Trivaldos for managing the Uranus cluster in Carlos III University where part of the simulations for this study were performed.
\hfill \break

\end{document}